\documentclass{appolb}
\usepackage{epsfig}

\def\csy{c_{\rm sym}}
\def\lf{\left} 
\def\ri{\right}
\def\Ksy{K_{\rm sym}}
\def\asya{a_{\rm sym}(A)}

\begin{document}
\title{Nuclear symmetry energy and neutron skin thickness%
\thanks{Presented at XXXII Mazurian Lakes Conference on Physics, Piaski, 2011}%
}
\author{M. Warda
\address{Katedra Fizyki Teoretycznej, Uniwersytet Marii Curie--Sk\l odowskiej,\\
        ul. Radziszewskiego 10, 20-031 Lublin, Poland}
\and
M. Centelles, X. Vi\~nas
\address{Departament d'Estructura i Constituents de la Mat\`eria \\
and Institut de Ci\`encies del Cosmos,\\
Facultat de F\'{\i}sica, Universitat de Barcelona,\\
Diagonal 647,  08028 Barcelona, Spain}
\and
X. Roca-Maza
\address{INFN, sezione di Milano, via Celoria 16, I-20133 Milano, Italy
}
}
\maketitle
\begin{abstract}
The relation between  the slope of the nuclear symmetry energy at saturation density and the neutron skin thickness is investigated.
Constraints on the slope of the symmetry energy are deduced from the neutron
skin data obtained in  experiments with antiprotonic atoms. Two types of neutron
skin are distinguished: the "surface" and the "bulk".  A combination of both types
forms neutron skin in most of nuclei. A prescription to calculate neutron skin
thickness and the slope of symmetry energy parameter $L$ from the parity
violating asymmetry measured in the PREX experiment is proposed.
\end{abstract}
\section{Introduction}
The symmetry energy is a quantity of a great importance in nuclear physics and
astrophysics \cite{bao08,ste05, lat07,bar05,hor01}. It governs numerous isospin-dependent properties of nuclei such as the
binding energy, the location of the drip lines, the density distributions,
as well as the reactions: giant resonances, heavy ion collisions, isospin
diffusion, and multifragmentation. At the same time, the nuclear symmetry
energy is crucial in the astrophysical calculations 
of neutron stars, supernova explosions and stellar nucleosynthesis.

Despite of its importance in nuclear physics, the symmetry energy is hardly known at densities different from saturation \cite{pie07,she07}. Nuclear models predict a large variety of its density dependence.
The values of the important parameter $L$, which describes the slope of the symmetry
energy at saturation, are spanned over a wide range. The so called "soft"
symmetry energy can be found for the D1S parametrization where $L=22$ MeV
whereas $L= 118$ MeV is calculated for the NL3 parameter set which predicts
"stiff" symmetry energy. To fix the symmetry energy at subsaturation
densities is needed for a better description of the isospin dependence of
nuclear interactions.

The neutron skin thickness is one of the observables where symmetry energy shows up
in the ground state of nuclei \cite{bro00}. Neutron skin defined through the rms radii of
protons and neutrons depends on the properties of the nuclear surface.
The relative differences of the neutron and the proton distributions in this
region are sensitive to the symmetry energy at the subsaturation densities. Thus
important information about nuclear forces can be deduced from the analysis of
the neutron skin properties.

In this paper we will study how neutron skin depends on the slope of the
symmetry energy at saturation by means of the droplet model (DM) of nuclei. The
predictions of the parameter $L$ of symmetry energy will be made based on the
neutron skin measurements in the antiprotonic atoms. The non-negligible
influence of the differences of the diffuseness of the neutron and the proton
density distribution at the nuclear surface on the value of neutron skin
thickness will be shown. The relations between the neutron skin
thickness of $^{208}$Pb and the slope of symmetry energy parameter $L$ with the
parity violating asymmetry measured in the PREX experiment \cite{prex1} will be investigated.

\section{Determination of the slope of symmetry energy}

\begin{figure}[t]
\begin{center}
\includegraphics[width=0.80\textwidth,angle=270,clip=true]{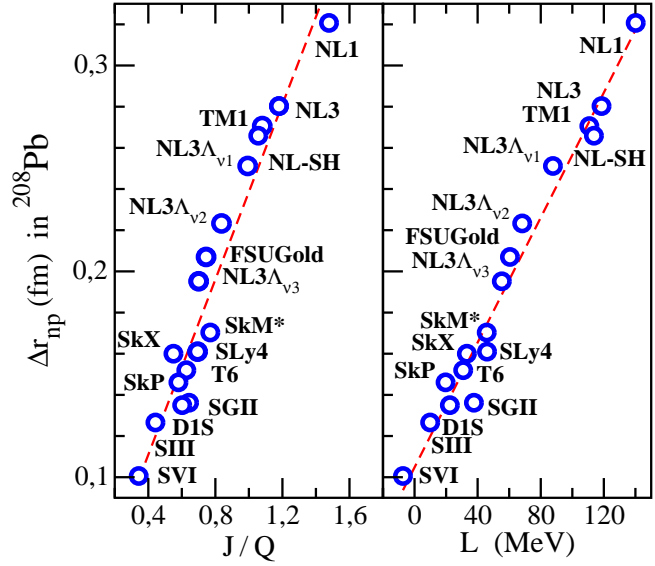}
\caption{(Color online) The neutron skin thickness in $^{208}$Pb calculated in various models plotted as a function of the ratio $J/Q$ of the symmetry energy at saturation to the surface stiffness coefficient (left) and the slope of symmetry energy coefficient $L$ (right).}
\end{center}
\end{figure}

The energy per particle of asymmetric nuclear matter can be expressed as:
\begin{equation}
e(\rho,\delta) = e(\rho,0) + \csy(\rho) \delta^2 + {\cal O}(\delta^4) \;,
\end{equation}
where  $\delta=(\rho_n-\rho_p)/\rho$ is the asymmetry between the densities of protons and neutrons.
This formula defines the symmetry energy
of a nuclear EOS $\csy(\rho)$, which can be  expanded around saturation in the form:
\begin{equation}
\csy(\rho) =  J - L \epsilon + \half \Ksy \epsilon^2 + {\cal O}(\epsilon^3)\;,
\end{equation}
where $\epsilon=(\rho_0-\rho)/(3\rho_0)$.
While the symmetry energy at saturation $J= \csy(\rho_0)$ is constrained by
the empirical information to be around 32 MeV \cite{vre03,car10,zen10,che05},
the predictions for the slope $L=3\rho\partial\csy(\rho)/ \partial\rho |_{\rho_0}$ and the curvature $\Ksy= 9\rho^2\partial^2 \csy(\rho)/ \partial\rho^2 |_{\rho_0}$ parameters are varying substantially among the different theoretical models \cite{war09}.

The neutron skin thickness defined as the difference of the neutron and the proton rms radii
\begin{equation}
\Delta r_{np} \equiv \langle r^2 \rangle_n^{1/2} 
- \langle r^2 \rangle_p^{1/2} 
\label{rnp}
\end {equation}
is described in the droplet model \cite{mye69}  by the formula
\begin{equation}
\Delta  r_{np}=\sqrt{\frac{3}{5}}\left[t- \frac{e^2 Z}{70J}
+\frac{5}{2R}\left(b_n^2-b_p^2\right) \right] \;.
\end {equation}
The second term in this expression is due to the Coulomb repulsion of protons
and the last term depends on the surface diffuseness of neutrons $b_n$ and
protons $b_p$. If as in the original papers of the droplet model $b_n$ and $b_p$
are assumed to be equal this term vanishes.
The leading term in Eq. (4) is almost a linear function of relative neutron
excess $I=(N-Z)/A$:
\begin{equation}
t=\frac{3}{2} r_0 \, \frac{J}{Q}\;
\frac{\displaystyle I-I_C}
{\displaystyle 1+(9J/4Q)A^{-1/3}} \;,
\end {equation}
where $I_C=\frac{e^2Z}{20Jr_0A^{1/3}}$.

The droplet  model shows a linear correlation of neutron skin in any heavy
nucleus with the ratio $J/Q$ of the symmetry energy at saturation to the surface
stiffness coefficient. Such correlation for the case of $^{208}$Pb calculated in
several nuclear forces is plotted in the left panel of Fig. 1. More surprising
is that the neutron skin in $^{208}$Pb is also linearly correlated with the
parameter $L$ what has been observed previously \cite{bro00,typ01} (see the
right panel of Fig. 1).

To explain this relation, first, we transform the expression (5) into the form
\begin{equation}
t=\frac{2r_0}{3J} \, \lf[J-\asya\ri] A^{1/3} \, (I-I_{\rm C})  \;,
\end {equation}
where $\asya= J/(1+(9J/4Q)A^{-1/3})$ is the symmetry energy coefficient
of the finite nucleus in the
droplet model. We have shown that in the heavy nuclei with mass
$A\ge 200$, $\asya$ is equal to $\csy(\rho)$ at densities around 0.1
fm$^{-3}$ \cite{cen09}. This relation is fulfilled in good approximation
also for lighter isotopes if $\csy(\rho)$ is taken at a slightly smaller subsaturation
density $\rho_A$ \cite{cen09}. Within the mean-field approach, such density can be parametrized
in terms of the mass number of the finite nucleus as follows:
$ \rho_A=\rho_0-\rho_0/(1+c A^{1/3})$, with the constant $c$ fitted to reproduce the
density of 0.1 fm$^{-3}$ for  $^{208}$Pb [see Ref. \cite{cen09} for further details].
In this way, $\asya$ in Eq. (6) can be replaced by $\csy(\rho_A)$ and
using Eq. (2) we can obtain
\begin{equation}
t= \frac{2r_0}{3J} \, L \,
 \Big(1 - \epsilon \frac{\Ksy}{2L}\Big) \epsilon A^{1/3} 
 \big(I-I_{\rm C}\big) .
\end {equation}
In the last formula the leading term is $t= \frac{2r_0}{3J} \, L \,I$.
As mentioned before,
the coefficient $J=c_{\rm sym}(\rho_0)$ of the symmetry energy
at saturation density has been empirically constrained to be around
32 MeV \cite{vre03,car10,zen10,che05}, which is well reproduced by the successful
nuclear forces \cite{war09}. However, the values of the parameter $L$ of the slope
of symmetry energy at saturation are varying more substantially
among the different theoretical models \cite{war09,cen09,cen10,roc11}.
As $J$ varies in a much narrower range than $L$ in the nuclear models, and as
if $J$ is larger (smaller) also $L$ is usually larger (smaller), Eq. (7) suggests
to leading order a linear correlation between the neutron skin thickness
in $^{208}$Pb and the parameter $L$, that is confirmed by the theoretical calculations
as it can be seen in Fig. 1.

From the independent calculations of $b_n$ and $b_p$ in the semi-infinite
nuclear matter \cite{war09,cen09} the influence of the diffuseness term in Eq.
(4) on the total neutron skin thickness can be estimated. In the case of 
$^{208}$Pb we have obtained values from around 0.035 fm to around 0.055 fm in
various nuclear models. These numbers mean that diffuseness term can contribute
as much as 20\% to 40\% to the total neutron skin thickness and they are too
large to be omitted. Moreover, the neutron skin thickness calculated in the
droplet model including the diffuseness term is very close to extended
Thomas-Fermi predictions, whereas without this term the two models differ 
significantly. These results show that the assumption of equal surface
diffuseness of neutrons and protons is not consistent with mean-field
predictions of density distribution in a nucleus.

The experimental neutron skins deduced from the antiprotonic atoms
\cite{trz01,jas04} presented in Fig. 2 show a linear increase of neutron skin
with $I$. This trend is consistent with the droplet model prediction in Eq. (4).  Since
we have shown that the slope of the neutron skin with $I$ depends
on the parameter $L$ (see Eq. (7)), the value of this parameter may be
constrained by experimental
data. The least squares fit of the formula (4) to the experimental data from the
antiprotonic atoms, also
presented in Fig. 2, reproduces average increasing trend of measurements. From
the fitting procedure including the diffuseness term, we
have found the value of $L=55 \pm 25$ MeV for the slope of symmetry energy,
which points towards a "soft" nuclear symmetry energy. The uncertainty of the value
of $L$ comes from large experimental errorbars of the measurements.

\begin{figure}[t]
\begin{center}
\includegraphics[width=0.60\textwidth,angle=-90,clip=true]{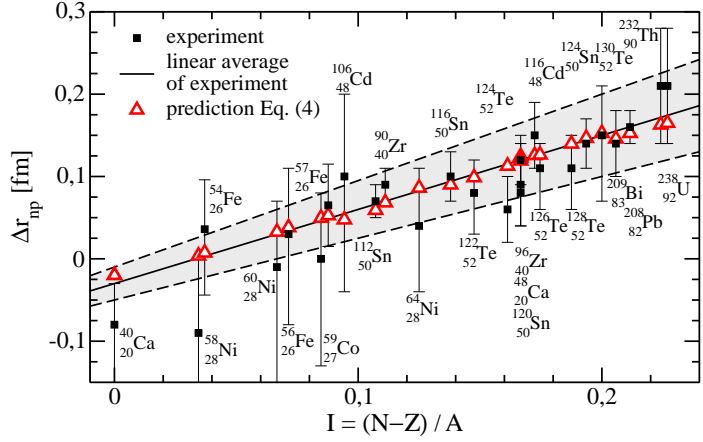}
\caption{(Color online) Neutron skin measurements in antiprotonic atoms\cite{trz01,jas04}  as a
function of neutron excess $I$. Average line of the experimental data is also
given as well as results of the fit of formula (4) to the average line.}
\end{center}
\end{figure}

\section{The "bulk" and the "surface" types of neutron skin}
\begin{figure}[t]
\begin{center}
\includegraphics[width=0.30\textwidth,angle=270]{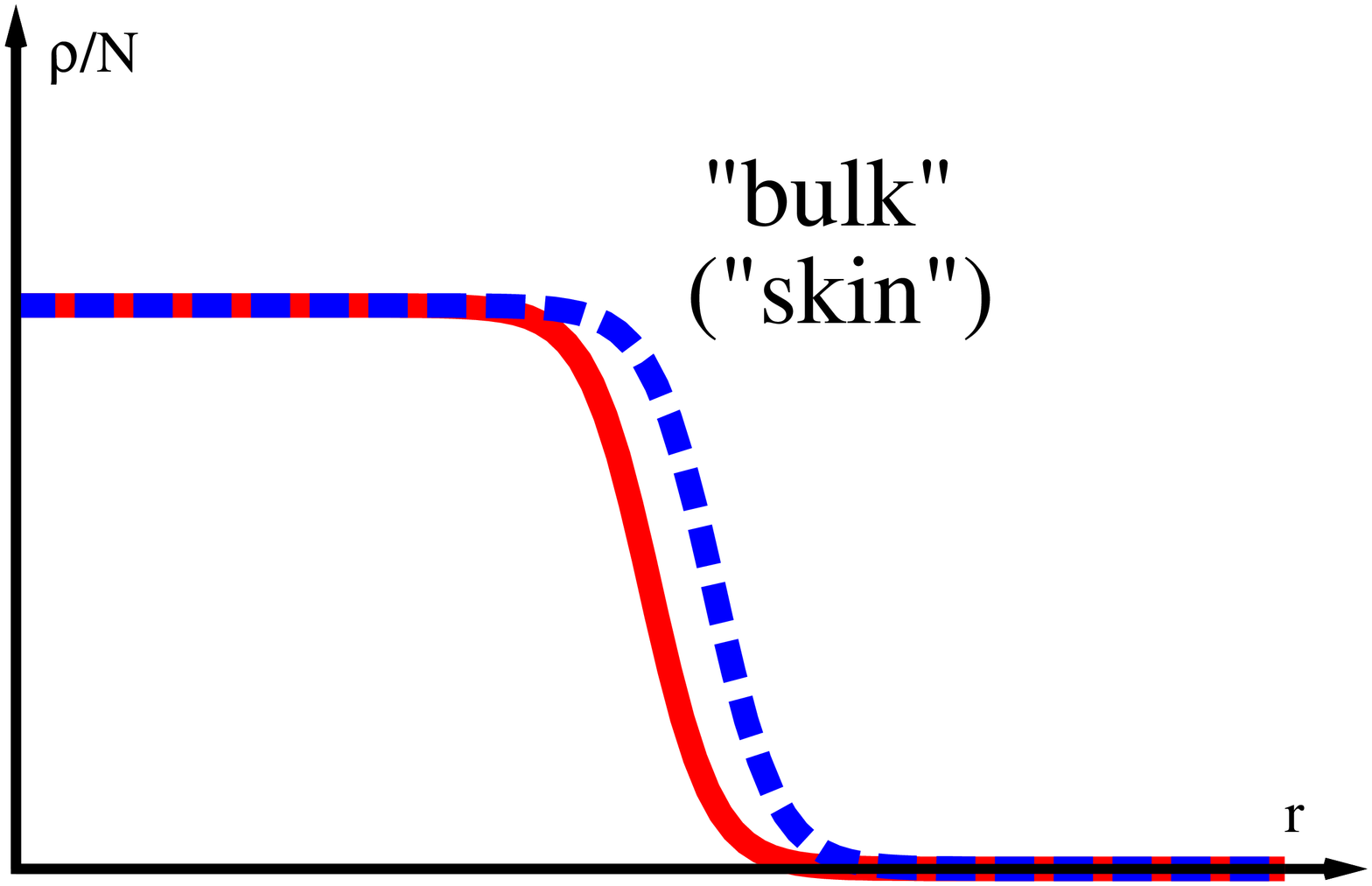}
\includegraphics[width=0.30\textwidth,angle=270]{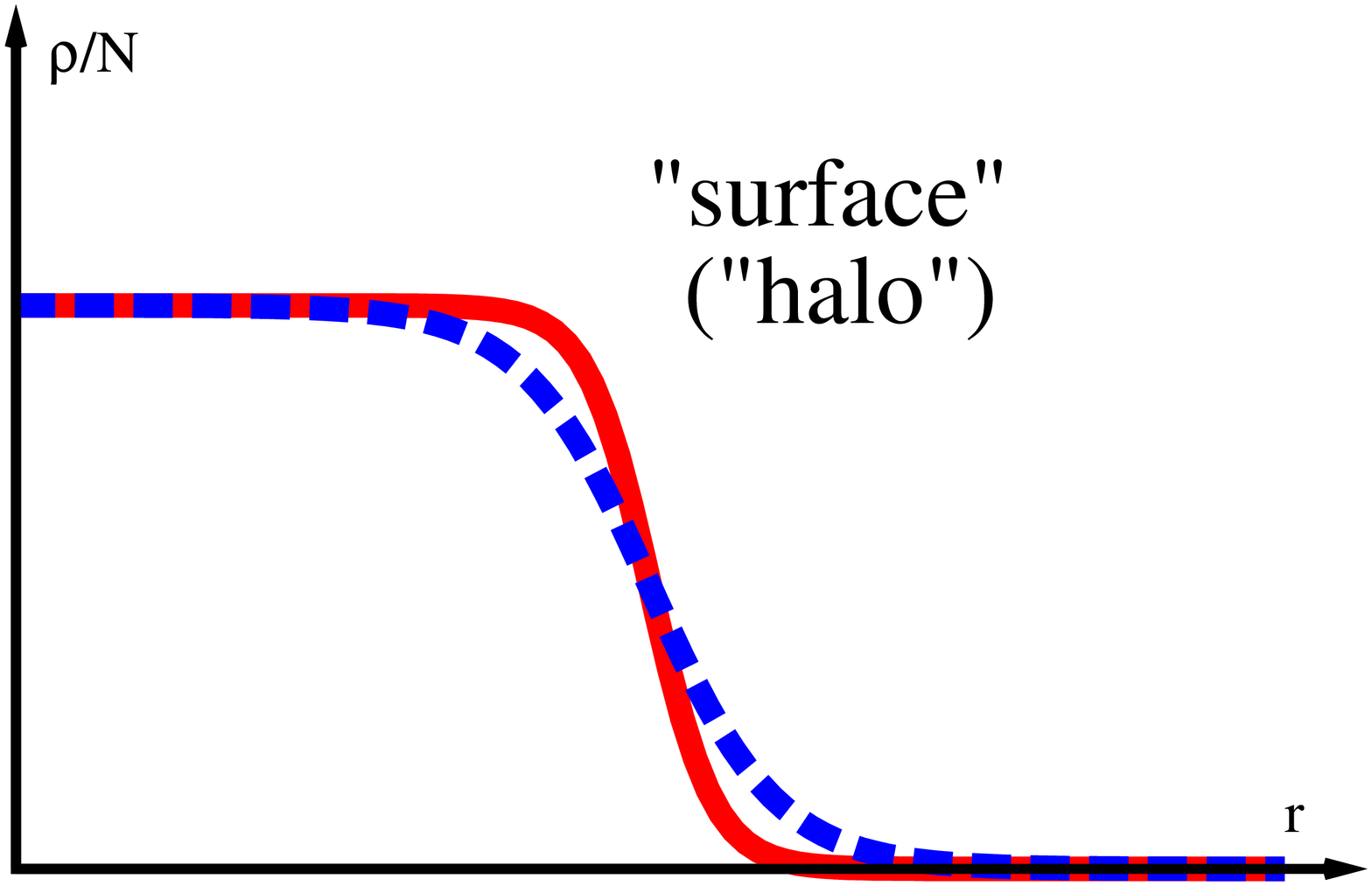}
\caption{(Color online) Schematic view of the "bulk" ("skin") and the "surface" ("halo") types of neutron skin.}
\end{center}
\end{figure}

\begin{figure}[t]
\begin{center}
\includegraphics[width=0.80\textwidth,angle=0,clip=true]{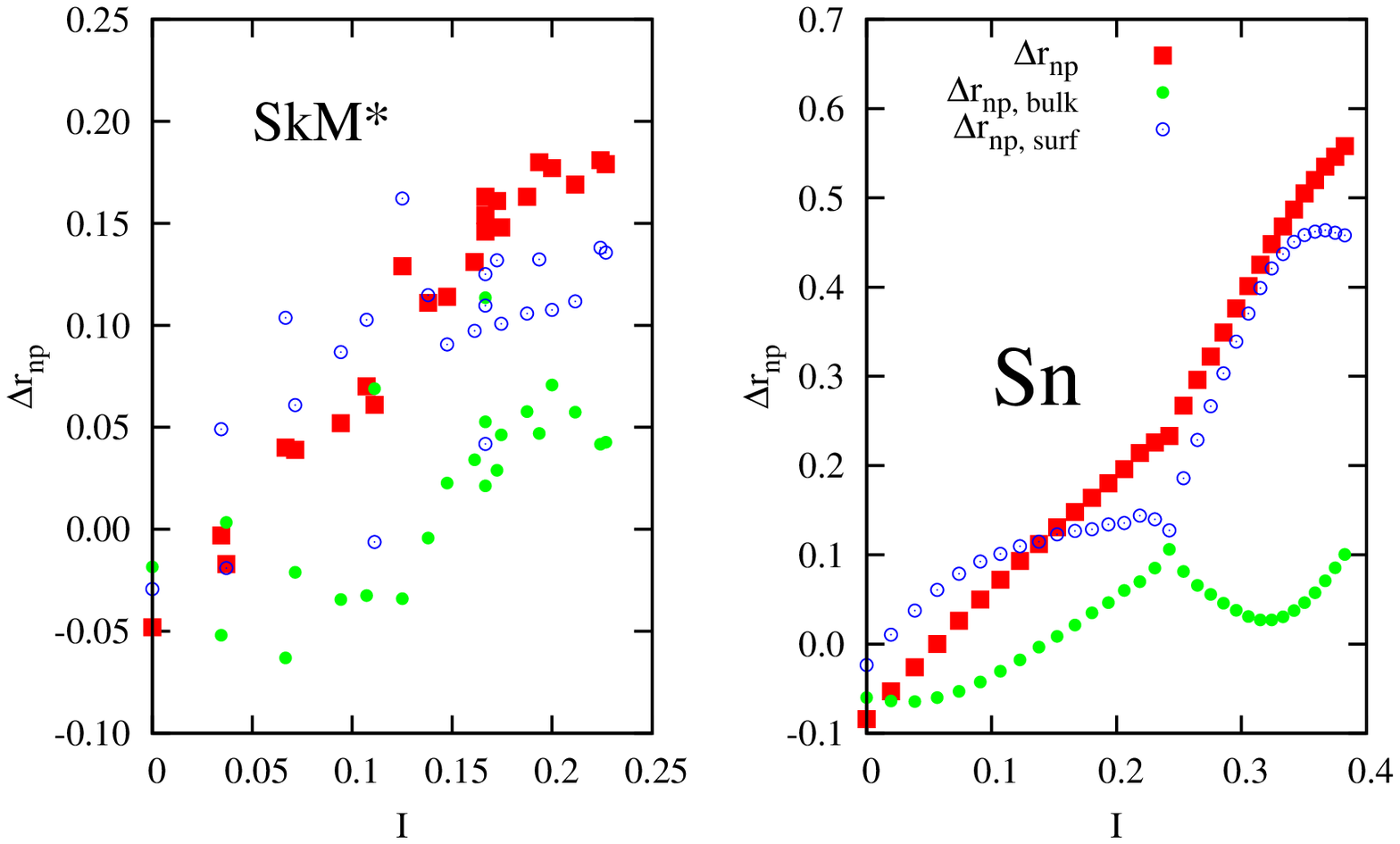}
\caption{(Color online) Neutron skin calculated with the SkM* Skyrme force in the antiprotonic atoms (left) and in the Sn isotopic chain (right).  The "bulk" and the "surface" contributions are also shown. 
 }
\end{center}
\end{figure}

Two basic scenarios leading to create neutrons skin can be distinguished. They
are schematically shown in Fig. 3. In the first type, called "skin" mode,  the
neutron surface is shifted outwards the proton surface. The alternative scenario is an
increasing of the diffuseness of the neutron surface relative to the proton one.
It is called the "halo" mode. When two parameter Fermi (2pF) function
\begin{equation}
\rho(r)=\frac{\rho_0}{1+ \exp{ [(r-C)/a] }} 
\end{equation}
is assumed for the neutron and the proton density distributions, both types of
neutron skin are easy to classify. The "skin" mode corresponds to 2pF
parametrization with $a_n=a_p$, whereas the pure "halo" mode is obtained with
$C_n=C_p$.
 The central density is not properly reproduced by the sharp-edge sphere of
a radius $C$ with the volume conservation condition \cite{war10}.
Therefore, the aforementioned "skin" and "halo" definitions do not correspond to
the droplet model expression for the neutron skin thickness and cannot be
directly related to the particular terms in the neutron skin formula in Eq. (4).
Instead, we propose to use the "bulk" and the "surface" contributions
\cite{cen10,war10}. They are based on the same idea as the "skin" and the "halo"
types but with special attention paid on preservation of the properties of the
interior of a nucleus by the "bulk" part. This contribution is defined through
the equivalent sharp-edge radii \cite{has88} of neutrons $R_n$ and protons $R_p$
\begin{equation}
\Delta r_{np}^{\rm bulk} \equiv
\sqrt{\frac{3}{5}}\left(R_n-R_p\right) \simeq \sqrt{\frac{3}{5}}\left[(C_n -C_p)
+\frac{\pi^2}{3}
\left(\frac {a_n^2}{C_n}-\frac {a_p^2}{C_p}\right) \right] 
\end{equation}
The remaining part of neutron skin thickness states the  "surface" contribution.  
\begin{equation}
\Delta r_{np}^{\rm surf} \equiv\Delta r_{np}-\Delta r_{np}^{\rm bulk}
\simeq
\sqrt{\frac{3}{5}} \, \frac{5}{2}
   \Big(\frac{b_n^2}{R_n}-\frac{b_p^2}{R_p}\Big)
   \simeq \sqrt{\frac{3}{5}} \, \frac{5\pi^2}{6}
\left(\frac{a_n^2}{C_n}-\frac{a_p^2}{C_p}\right) \;.
\end{equation}
As it is clear from Eqs. (9) and (10), the "bulk" and the "surface" contributions
can be also easily expressed by the 2pF parameters.

In Fig. 4 the neutron skin thickness calculated with the SkM* parameter set is
plotted for the nuclei investigated in the antiprotonic atom measurements as a
function of neutron excess $I$. A linear increasing trend of neutron skin
thickness can be easily noticed. The "bulk" and the "surface" contributions
calculated for the same nuclei are also plotted in Fig. 4. To obtain these
values the mean-field density distributions have been parametrized by the 2pF
profile. Special care was taken on the precise reproduction of the surface
region. The "bulk" contribution roughly shows a linear increasing trend as the
total neutron skin with a similar slope. In the light nuclei with small $I$ this
contribution is negative due to the dominant influence of the Coulomb repulsion.
In the heavy nuclei it grows up and states roughly one third of the total value
of neutron skin. The "surface" contribution is very scattered especially in the
light nuclei. In the region of the large neutron excess this part remains at the
same level and does not change significantly with $I$. It seems that
single-particle effects in particular nucleus have an important influence on the
surface region of some nuclei what impacts strongly on the "surface" part of
neutron skin. On the other side, the "bulk" contribution is governed by the
classical liquid drop properties of nuclei and the influence of quantal effects
is limited.

In the Sn isotopic chain the single-particle influence on the nuclear density
distribution is limited to shell structure of neutrons. The increasing trend of
neutron skin as well as both "bulk" and "surface" contributions can be noticed.
 Again neutron
skin thickness is composed of the both contributions. Shell effects are pronounced at magic numbers ($I=0$ and $0.24$). 
A huge increase of the "surface" part can be noticed in the mid-shell nuclei whereas there is a kink with a local minimum for doubly magic nuclei. The "bulk" contribution shows up opposite and less pronounced  behaviour.
The "surface" part is
larger than the "bulk" one in most of the nuclei of the tin chain.

\section{Predictions of PREX experiment}
\begin{figure}[t]
\begin{center}
\includegraphics[width=0.80\textwidth,angle=270,clip=true]{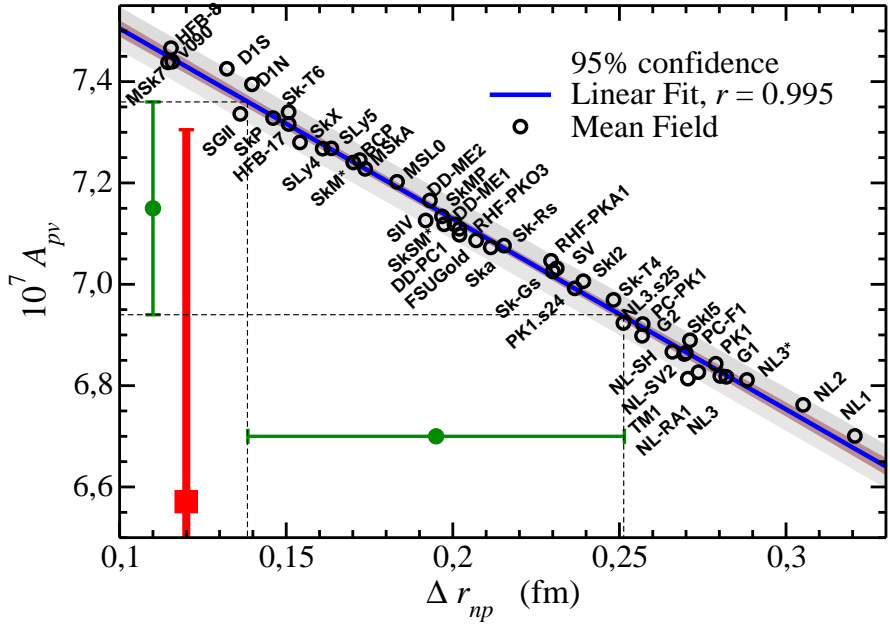}
\caption{(Color online) Relation between $A_{pv}$ and $ \Delta r_{np}$
calculated in various models. Predicted errorbars with arbitrary chosen central
value is plotted with thin green errorbar and preliminary measured value  \cite{prex1} with
thick red errorbar.}
\end{center}
\end{figure}

The PREX (Pb Radius Experiment) performed at Jefferson Lab aims to constraint the neutron radius of $^{208}$Pb by means of parity violating electron scattering (PVES) \cite{prex1,prex4,prex2,prex3}. In this experiment polarized electrons are scattered on the $^{208}$Pb target. Electrons couple with protons mainly via the electromagnetic ($\gamma$) interaction and with neutrons via the weak interaction ($Z^0$). For ultra-relativistic electrons, as in PREX conditions, the weak potential has opposite sign depending on the polarization (helicity) of the electron beam. Therefore, the elastic differential crosssection of right-handed ($d\sigma_+/d\Omega$) electrons is different to the one measured for left-handed ($d\sigma_-/d\Omega$) electrons. The difference of these cross sections allows one to define the parity violating asymmetry:
\begin{equation}
A_{pv}\equiv
\frac{\displaystyle \frac{d\sigma_+}{d\Omega}-\frac{d\sigma_-}{d\Omega}}
{\displaystyle \frac{d\sigma_+}{d\Omega}+\frac{d\sigma_-}{d\Omega}} \;.
\label{apv}
\end{equation}
It is measured at PREX at the fixed energy of $1.06$ GeV and the scattering angle of about $5^o$ ($q_{\rm lab}\approx 0.47$ fm${}^{-1}$). 
We calculate the elastic differential cross sections of ${}^{208}$Pb for a
polarized electron beam by means of the distorted
wave Born approximation (DWBA) \cite{roc11}. For that we have modified the code
used in Ref.~\cite{roc08}. In short, the DWBA consist on the
exact phase shift analysis of the Dirac equation, 
\begin{equation}
[\alpha{\bf p} + \beta m_e + V(r)]\psi_e = E\psi_e
\label{dirac}
\end{equation}
where $\psi_e$ is the electron wave function and $V(r) = V_{C}(r) \pm V_{W}(r)$ is the electroweak potential felt by the ultra-relativistic electron beam depending on their helicity. The connection between the parity violating asymmetry and the neutron skin thickness of a nucleus can be qualitatively understood from the analytical expression of the former within the plane wave Born approximation (PWBA),
\begin{equation}
A_{pv}^{\rm PWBA} = \frac{G_F q^2}{4\pi \alpha \sqrt{2}}
\Big[ 4 \sin^2\theta_W + \frac{F_n(q) - F_p(q)}{F_p(q)} \Big] ,
\end{equation}
where the neutron and proton formfactors $F_n(q)$ and $F_p(q)$, respectively,
are also written in PWBA. Therefore, in the low-momentum transfer regime,
$F(q)\approx 1 - q^2\langle r^2\rangle/6$ and $A_{pv}^{\rm PWBA}$ explicitly
depend on the factor $(\langle r^2\rangle_n^{1/2}-\langle
r^2\rangle_p^{1/2})\times(\langle
r^2\rangle_n^{1/2}+\langle r^2\rangle_p^{1/2})$. For realistic results,
however, one cannot use the simple PWBA and full DWBA calculations must be
performed \cite{prex1,prex4}.

In Fig. 5 the parity violating asymmetry in $^{208}$Pb calculated in DWBA for a
large sample of successful nuclear forces is plotted versus the neutron skin
thickness calculated for the same parametrizations. Models that fail to predict
the well-known charge radius and binding energy of $^{208}$Pb are not allowed in
Fig. 5. A very good linear correspondence of the parity violating asymmetry
and the neutron skin thickness is found. From the measurement of the
parity violating asymmetry, the neutron skin can be easily deduced using linear
dependence of Fig. 5:
\begin{equation}
A_{pv}\mathrm{(ppm)}= 0.788 - 0.375 \Delta r_{np}\mathrm{(fm)}
\end{equation}

From this relation, taking into account that all models properly reproduce proton rms radius, one can notice that the expected 3\% errorbars of $A_{pv}$ should allow to obtain the neutron radius of $^{208}$Pb with  1\% uncertainty. It is illustrated in Fig. 5 with the green errobars plotted at the arbitrary chosen value of $A_{pv}$. The PREX experiment completed a successful run in 2010.
The statistical error of the first run of the experiment was larger than 3\% \cite{prex1}
\begin{equation}
A_{pv}^{\rm exp}=0.6571 \pm0.0604 (stat) \pm 0.0130 (syst)
\end{equation}
This preliminary value of $A_{pv}$ with the reached 10\% accuracy, marked in Fig. 5 by the thick red errorbar, does not  indicate on the forces 
which properly describe both the parity violating asymmetry as well as neutron skin.
The PREX-II proposal aims to improve the accuracy of the experiment
\cite{prex1}.

The slope of symmetry energy parameter $L$ can be also calculated for each force
presented in Fig. 5. As $A_{pv}$ is linear with  $\Delta r_{np}$ and $\Delta
r_{np}$ is linear with $L$ (see Fig. 1) we may also expect a linear relation
between $A_{pv}$ and $L$.
From the confrontation of these two quantities we can find a nice correlation
with average fitted line \cite{roc11}
\begin{equation}
A_{pv}\mathrm{(ppm)}=0.750 - 0.000557 L\mathrm{(MeV) }
\end{equation}
The PREX experiment can provide information not only about the neutron skin in
$^{208}$Pb but also the independent estimation of $L$.

\section{Conclusions}
We have discussed  various aspects of neutron skin and its relations to the symmetry energy of nuclear matter. The linear relation between the thickness of neutron skin and the slope of the symmetry energy coefficient $L$ has been proven. With the use of this correlation  we have deduced from the neutron skin measurements in the antiprotonic atoms relatively small value of $L=55$ MeV which hints the soft character of symmetry energy.

Neutron skin can be produced as a result of the shift of the neutron surface in relation to the proton one (the "bulk" type), or due to the differences in the surface thickness of proton and neutron matter in a nucleus (the "surface" type). In the general case both mechanisms contribute to the creation of neutron skin. 

The precise prescription for calculating the neutron skin thickness in $^{208}$Pb and the parameter $L$ form the parity violating asymmetry in the PREX experiment has been given. High linear correlation between  $A_{pv}$ and $\Delta r_{np}$ in $^{208}$Pb  has been noticed.

\section{Acknowledgements}
 Work partially supported by the 
Spanish Consolider-Ingenio 2010 Programme CPAN CSD2007-00042 and by Grants No.\ 
FIS2008-01661 from MICINN (Spain) and FEDER, No.\ 2009SGR-1289 from Generalitat de
Catalunya (Spain), and No. DEC-2011/01/B/ST2/03667 from NCN (Poland). X.R. acknowledges grant from INFN (Italy).


\end{document}